\documentclass[a4paper,aps,prb,reprint,groupedaddress,showpacs,showkeys,nofootinbib]{revtex4-1}
\usepackage[dvips]{graphicx}
\usepackage{amsmath}
\usepackage{array}
\usepackage{bm}
\usepackage{fancyhdr}
\usepackage{makeidx}
\usepackage{epsfig}
\usepackage{amsfonts}
\usepackage{amssymb}
\usepackage{color}

\begin{document}

\title{Probing the wavefunction of the surface states in Bi$_2$Se$_3$
topological insulator: a realistic tight-binding approach} \author{A.~Pertsova
and C.~M.~Canali} \affiliation{Department of Physics and Electrical Engineering,
Linn{\ae}us University, Norra V{\"{a}}gen 49, 391 82, Kalmar, Sweden}
\date{\today}

\begin{abstract} 
We report on microscopic tight-binding modeling of surface
states in Bi$_2$Se$_3$ three-dimensional topological insulator,  based  on a
\textit{sp}$^3$ Slater-Koster Hamiltonian, with parameters calculated from
density functional theory.  The effect of spin-orbit interaction on the
electronic structure of the bulk and of a slab with finite thickness is
investigated.  In particular, a phenomenological criterion of band inversion is
formulated for both bulk and slab, based on the calculated atomic- and
orbital-projections of the wavefunctions, associated with valence and conduction
band extrema at the center of the Brillouin zone.  We carry out a
thorough analysis of the calculated bandstructures of slabs with varying
thickness, where surface states are identified using a quantitative criterion
according to their spatial distribution. The thickness-dependent energy gap,
attributed to inter-surface interaction, and the emergence of gapless surface
states for slabs above a critical thickness are investigated. We map out the
transition to the infinite-thickness limit by calculating explicitly the
modifications in the spatial distribution and spin-character of the surface
states wavefunction with increasing the slab thickness.  Our numerical analysis
shows that the system must be approximately forty quintuple-layers thick to
exhibit completely decoupled surface states, localized on the opposite
surfaces. These results have implications on the effect of external perturbations
on the surface states near the Dirac point.

\end{abstract}
 
 \maketitle

\section{Introduction}\label{intro}

Topological insulator~\cite{Hasan,XLQi} (TI) materials host on their boundaries
a novel type of topological states of quantum matter, which, unlike the quantum
Hall state, exist without the breaking of time-reversal
symmetry.~\cite{KaneMele1,KaneMele2} Theoretical prediction and subsequent
experimental demonstration of these topological states in both two-
\cite{Bernevig, Konig} (2D) and
three-dimensional~\cite{Fu2007,FuKane,Moore2007,Roy,Teo2008,Zhang2009,HZhang,
Hsieh2008,Xia,Hsieh2009,Chen} 
(3D) systems have given rise to what is now one of the most rapidly developing
fields in condensed matter physics. Apart from providing a test platform for
fundamental concepts, the study of TIs holds promise for novel  applications in
materials science and chemistry,~\cite{Kong} spintronics~\cite{Pesin} and
quantum computation.~\cite{Read,Nayak} However, to be able to fully explore the
potential of TIs, it is essential to have a detailed knowledge of the nature and
properties of topological surface states in real TI materials,~\cite{Cao} as
well as a quantitative understanding of  how they respond to external
perturbations.~\cite{QLiu,Wray,Beidenkopf}  Experimentally, these questions are
being addressed with advanced surface-sensitive experimental probes, such as
spin- and angle-resolved photoemission spectroscopy~\cite{Hsieh2009}
[(SR)-ARPES] and scanning tunneling microscopy~\cite{Roushan,Hor} (STM). 

Along with experimental advances, there is a growing need  for atomistic
modeling of TIs that would enable quantitative predictions and direct comparison
with experiment. Significant progress has been made in using \textit{ab initio}
methods to calculate electronic~\cite{HZhang,WZhang,KPark,Zhao} and
magnetic~\cite{Niu,JMZhang,Henk,Abdalla} properties of TIs.  However, such
methods suffer from severe computational limitations, particularly in the case
of slab geometry as well as surface supercell calculations, which are employed
in studies of impurity-doping effects. In addition, more accurate \textit{ab
initio} methods often lack the conceptual  transparency and flexibility of the
model Hamiltonian approaches, which have been of fundamental importance for
driving progress in this research field.~\cite{KaneMele1,KaneMele2}  Microscopic
tight-binding (TB) models, which have already proved successful in quantitative
description of electronic and magnetic properties of
semiconductors,~\cite{Tang,Strandberg} may provide a convenient platform to
address similar issues in TIs. Several studies have recently appeared in the
literature, in which TB descriptions with different level of complexity have
been introduced, ranging from  models built on a simplified lattice
structure~\cite{BlackSchaffer} or a restricted orbital basis set inferred from
symmetry arguments~\cite{Mao,CXLiu} to fully microscopic models, with parameters
extracted from density functional theory
(DFT).~\cite{Kobayashi,WZhang,Bahramy,Barfuss} To date, the latter class of
models is still the least represented among the model Hamiltonian approaches to
TIs.

In this work we employ a microscopic TB model to study the properties of surface
states in Bi$_2$Se$_3$, a prototypical 3D TI, which belongs, along with
Bi$_2$Te$_3$ and Sb$_2$Te$_3$,  to the family of binary tetradymite
semiconductors with layered structure.~\cite{Landolt} Although these materials
have been studied for decades due to their excellent thermoelectric
properties,~\cite{Misra,Urazhdin1,Urazhdin2} they have recently attracted
considerable attention as 3D TIs, e.g. materials that exhibit topologically
protected conducting surface states with linear (Dirac) dispersion and helical
spin-texture, traversing the bulk insulating
gap.~\cite{Hsieh2008,Xia,Hsieh2009,Chen} Due to a relatively large band gap
(0.3~eV for Bi$_2$Se$_3$) and rather simple surface states, consisting of a
single Dirac cone,~\cite{Hsieh2009} the  Bi$_2$Se$_3$ family of 3D TIs  is the
most studied  both experimentally and theoretically. 

Our treatment is based on the \textit{sp}$^3$ Slater-Koster
Hamiltonian.~\cite{Koster} We use the parametrization developed by
Kobayashi,~\cite{Kobayashi} by fitting to DFT
calculations. Throughout  this work, our strategy has been to make use of the
computational efficiency and simplifications, offered by the TB approach, in
order to investigate key features of the surface states in Bi$_2$Se$_3$ 3D TI,
which are inaccessible by \textit{ab initio} methods. Importantly, we consider
slabs with thicknesses ranging  from 1 to 100 quintuple layers (QLs), which
corresponds to length scales in the range of 1-100 nm. In contrast, thicknesses
typically investigated in \textit{ab-initio}--based studies 
do not exceed several
quintuple layers.~\cite{Zhao} In agreement with previous
reports,~\cite{YZhang,Linder,Lu2010} we find a gap due to interaction between
opposite surfaces, which decreases with increasing the slab thickness. Starting
from 5 QLs, the size of the gap becomes smaller than $10^{-3}$~eV, and one can
identify surface states with linear dispersion and helical spin-texture. For
each slab thickness we determine the surface character of Bloch states using the
procedure put forward in Ref.~\onlinecite{KPark}, i.e. based on the
contribution of the real-space projected wavefunction onto the two surfaces of
the slab. 

Explicit calculations of the atomic- and orbital-projections of the
wavefunctions, associated with valence and conduction band extrema in both bulk
and slab geometry, allowed us to construct a phenomenological picture of band
inversion. The latter effect is induced by spin-orbit interaction and
is responsible for the occurrence of topological surface states across
the bulk insulating gap.~\cite{HZhang}  Furthermore, based on a similar
analysis, we were able to track the changes in  the spatial distribution and the
spin character of the surface states wavefunctions at and in the vicinity of the
Dirac point, for increasing slab thickness.  Our calculations showed that the
states corresponding to top and bottom surfaces become completely decoupled,
i.e. spatially separated,  only for very thick slabs 
containing $40$ QLs. We also
calculated the spin-orientation of the surface states in momentum space as a
function of thickness. The disturbances in the helical spin-texture, expressed
through in-plane and out-of-plane tilting angles of the spin, are shown to be
significant for thin slabs up to 10 QLs and are manifestations of both
inter-surface interaction and the proximity to bulk states.       

The rest of the paper is organized as follows. In Section~\ref{theo} we discuss
the details of the TB model and some computational aspects.  The results of the
simulations are presented in Section~\ref{results}. We begin in
Section~\ref{bulk} by analyzing the effect of SO on the bulk bandstructures,
where the bulk band inversion is interpreted as a characteristic change in the
spatial distributions of the orbital-resolved projections of the wavefunctions
at conductance and valence band extrema at the $\bar{\Gamma}$ point. The results
of surface bandstructure calculations are presented in Section~\ref{slab}, in
particular the spatial character of the states emerging within the bulk gap and
the opening of the gap at finite thicknesses are described quantitatively in
Section~\ref{gap}. We also discuss the procedure to identify the inverted
character on conduction and valence bands in the presence of SO for a slab
geometry.  In Section~\ref{wf_atG}, based on the analysis of the wavefunctions
of the states at the Dirac point, a quantitative criterion of an infinitely
thick slab, with no interaction between the surfaces, is formulated. In
addition, we comment on the orbital character of the wavefunctions at the Dirac
point, especially on the non-negligible contribution of the in-plane
$p$-orbitals. The spin-properties of the Dirac-cone states and deviations from
the perfect spin-momentum locking due to inter-surface interaction and the
presence of bulk states are calculated in Section~\ref{spin}. 
Finally, Section~\ref{concl} contains our conclusions. 

\section{Methods}\label{theo}

\begin{figure}[ht!]
\begin{center}\includegraphics[width=0.98\linewidth,clip=true]{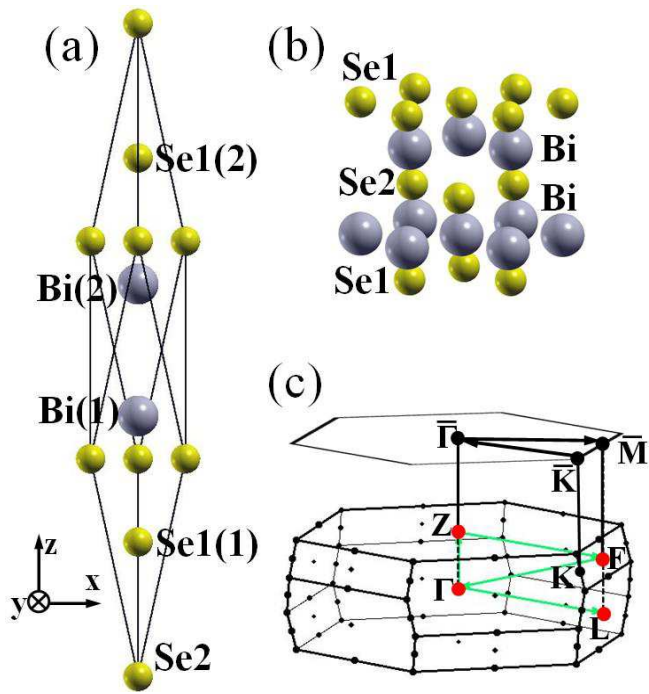}\end{center}
\caption{(Color online) Crystal structure of Bi$_2$Se$_3$. (a) Rhombohedral unit cell of the bulk crystal. 
Se1 and Se2 refer to two non-equivalent positions of Se atoms. 
(b) The structure of 1 QL formed by five alternating atomic layers of Bi and Se. 
(c) The Brillouin zone of the bulk crystal and that of the (111) surface of Bi$_2$Se$_3$.}\label{fig0} 
\end{figure}

We begin with a brief description of the crystal structure and chemical bonding
of materials belonging to the Bi$_2$Se$_3$ family. 
Bi$_2$Se$_3$ has a rhombohedral crystal
structure with  crystallographic $R\bar{3}m$ [$D^{5}_{3d}$] space group, with
five atoms in the unit cell [see Fig.~\ref{fig0}(a)].
The crystal is formed by
stacking of hexagonal monolayers of  either Bi or Se in a close-packed fcc (ABC)
fashion.  The sequence of five atomic layers, i.e. Se1-Bi-Se2-Bi-Se1, forms a QL
[see Fig.~\ref{fig0}(b)], with Se1 and Se2 indicating two nonequivalent
positions of Se atom. Hence it is convenient to describe the crystal structure
of bulk Bi$_2$Se$_3$ in terms of QLs stacked along the direction perpendicular
to atomic planes (\textit{c}-axis). 

The chemical bonding between atoms in the atomic layers within a QL is of
covalent-ionic type with dominant covalent character. 
Adjacent QLs are weakly bound
through Se1-Se1 bonds by van der Waals forces,~\cite{Misra} 
allowing easy cleavage
of Bi$_2$Se$_3$ on the (111) Se surface plane.  Based on the electronic
configurations of Bi and Se, with the outermost orbitals being of $p$-character
for both types of atoms, it is reasonable to expect that the bonding within QLs
is mainly due to interactions between $p$-orbitals. 
In fact, a simple but useful physical
picture of the chemical bonding present in QLs 
features  $pp\sigma$-chains formed by strongly 
interacting $p$-orbitals of atoms
in nearest-neighbors atomic layers. 
Evidence of this coupling can be indirectly inferred from 
STM topographies of Bi-antisites defects in Bi$_2$Se$_3$.~\cite{Urazhdin1}
In Sections~\ref{bulk} and \ref{slab}, we
will comment on the orbital character of the valence and conduction band extrema
in bulk Bi$_2$Se$_3$ and of surface states in Bi$_2$Se$_3$ slab, respectively.

For modeling of the electronic structure of Bi$_2$Se$_3$ we employ the
\textit{sp}$^3$ TB model with Slater-Koster parameters obtained by
Kobayashi~\cite{Kobayashi} by fitting to bulk bandstructures calculated with
DFT. Interactions between atoms in the same atomic layer and between atoms in
first and second nearest-neighbor layers are included. The spin-orbit
interaction (SO), which is the key element leading to a non-trivial bulk band
structure and metallic surface states, is incorporated in the intra-atomic
matrix elements.~\cite{Harrison} Thus the Hamiltonian of the system reads
\begin{eqnarray}\label{eq:1} \hat{H}(\bf{k}) & = & \sum_{\substack{ii^{\prime},\sigma\\
\alpha\alpha^{\prime}}} t_{\scriptsize ii^{\prime}}^{\scriptsize
\alpha\alpha^{\prime}}\,e^{i{\bf k}\cdot{\bf r}_{ii^{\prime}}}\,\hat{c}^{\sigma\dagger}_{i\alpha}
\,\hat{c}^{\sigma}_{i^{\prime}\alpha^{\prime}}\\\nonumber & + &
\sum_{\substack{i,\sigma\sigma^{\prime}\\\alpha\alpha^{\prime}}}
\lambda_{\scriptsize i} \left\langle i,\alpha,\sigma \right|
\hat{\vec{L}}\cdot\hat{\vec{S}}\left| i,\alpha^{\prime},\sigma^{\prime}
\right\rangle \,\hat{c}^{\sigma\dagger}_{i\alpha}
\,\hat{c}^{\sigma^{\prime}}_{i\alpha^{\prime}}\:, 
\end{eqnarray} 
where 
$\mathbf{k}$ is the reciprocal-lattice vector that spans the Brillouin zone, 
$i(i^{\prime})$ is the atomic index, $\alpha(\alpha^{\prime})$ labels atomic
orbitals $\left\{ s\right.$, $p_x$, $p_y$, $\left.p_z \right\}$, and 
$\sigma(\sigma^{\prime})$ denotes the spin. Here $i$ refers to the atomic positions in the unit cell, while 
$i^{\prime}$$\ne$$i$ runs over all neighbors of atom $i$, including atoms in the adject cells, with $\mathbf{r}_{i i^{\prime}}$ 
being the vector connecting the two atoms ($\mathbf{r}_{i i^{\prime}}$=$0$ for $i$=$i^{\prime}$). The coefficients $t_{\scriptsize i
i^{\prime}}^{\scriptsize \alpha\alpha^{\prime}}$ are the Slater-Koster 
parameters (for $i$=$i^{\prime}$ $t_{\scriptsize i
i^{\prime}}^{\scriptsize \alpha\alpha^{\prime}}\equiv t_{\scriptsize i
i}^{\scriptsize \alpha\alpha}$ give the on-site energies) and $\hat{c}_{i\alpha}^{\sigma\dagger}(\hat{c}_{i\alpha}^{\sigma})$
is the creation(annihilation) operator for an electron with spin $\sigma$ at the
atomic orbital $\alpha$ of site $i$. The second term in Eq.~(\ref{eq:1})
represents the on-site SO, where $\left|i,\alpha,\sigma\right\rangle$ are spin-
and orbital-resolved atomic orbitals, $\hat{\vec{L}}$ is the orbital angular
momentum operator and $\hat{\vec{S}}$ is the spin operator;
$\lambda_{\scriptsize i}$ is the SO strength. We refer to
Ref.~\onlinecite{Kobayashi} for the exact parametrization of $t_{\scriptsize i
i^{\prime}}^{\scriptsize \alpha\alpha^{\prime}}$ and $\lambda_{\scriptsize i}$. 

In bulk-bandstructure calculations we use the rhombohedral unit cell with five nonequivalent atoms, 
with the cell repeated periodically in $x$-, $y$-
and $z$-directions. In calculations involving the Bi$_2$Se$_3$ ($111$) surface
we consider a slab consisting of $N$ quintuple layers, or, equivalently, of $5N$
atomic layers.  
Since each atomic layer is an equilateral triangular lattice, 
it is sufficient to assign one atom per each layer, which gives a total of $5N$
atoms in the slab unit cell. 
The slab is finite along the $z$-direction 
(QL-stacking axis), with the unit cell repeated periodically in the $x$-$y$
plane.  Using this TB model we were able to compute bandstructures of slabs with
thicknesses up to $100$ QLs with a reasonable computational cost.

\section{Results and discussion}\label{results} 
\subsection{Band inversion in bulk Bi$_2$Se$_3$}\label{bulk}

The presence of gapless edge or surface states, robust against
time-reversal-invariant perturbations, distinguishes a topological insulating
phase from a trivial one. However, in order to elucidate the origin of the
topological order in existing TIs and to facilitate the search for new TI
materials, it is necessary to have a set of criteria that allow us to
differentiate between topologically trivial and non-trivial insulators, based on
the information about their bulk properties. Within the framework of topological
band theory,~\cite{KaneMele2,Fu2007,FuKane,Moore2007,Roy,Fu2006}
time-reversal-invariant insulators are classified according to a $Z_2$
topological invariant, assigned to their bandstructure. In 2D there is a single
$Z_2$ invariant, which distinguishes the quantum spin-Hall state (2D TI), as in
HgTe/CdTe quantum wells, from an ordinary insulator. In 3D a set of four $Z_2$
invariants leads to a classification based of three classes, namely strong TIs,
weak TIs and ordinary insulators. For systems with inversion symmetry, the $Z_2$
invariant is  determined by the product of the parity eigenvalues 
of occupied bands 
at the time-reversal-invariant momenta in the Brillouin zone.~\cite{FuKane} 
Using this scheme, a few strong TIs have been predicted, such as strained
$\alpha$-Sn and HgTe, Bi$_{1-x}$Sb$_x$ and Bi$_2$Se$_3$-like 3D TIs. 

The emergence of topological order can be understood using the concept of
\textit{band inversion}.~\cite{FuKane,HZhang,Yan,Zhu2012} Clearly, a necessary
condition for a TI is the existence of a non-trivial bulk band gap. In fact,
in the above mentioned materials, a topological phase transition, or
alternatively a change in the value of the $Z_2$ invariant which in turn
implies the existence of gapless states on the boundary, 
is accompanied by 
a visible and well-defined change in the bandstructure. This change occurs
precisely in the insulating gap and can be observed as a function of an external
or an intrinsic parameter. In HgTe, which is a zero-gap semiconductor but
acquires a gap due to external potential in a quantum well structure, the bands
with $s$- and $p$-character are inverted with respect to their usual sequence at
the $\Gamma$ point.~\cite{Bernevig} A similar mechanism is realized in strained
$\alpha$-Sn, which has recently been shown to exhibit a 3D TI phase, with
Dirac-like surface states emanating from the second lowest valence band across
the strained-induced gap.~\cite{Barfuss} In Bi$_{1-x}$Sb$_x$ alloy the inverted 
bandstructure is characterized by a change in the order of the bands with 
even ($L_s$) and odd ($L_a$) parity at the $L$ point of the Brillouin zone 
compared to pure Bi. This inversion is induced by Sb doping 
and is due to the 
strong TI character of the valence band of pure Sb.~\cite{FuKane} In
Bi$_2$Se$_3$ 3D TI a non-trivial bulk band gap owns its existence to SO, with
the parity of the valence and conduction band inverted at the $\Gamma$
point.~\cite{HZhang} Without SO  the material would be a trivial insulator.
However, in the presence of SO  the change in the parity of one of the occupied
bands leads to a change in the value of the $Z_2$ invariant, signaling a
topological phase transition.

\begin{figure}[ht!]
\begin{center}\includegraphics[width=0.98\linewidth,angle=0,clip=true]{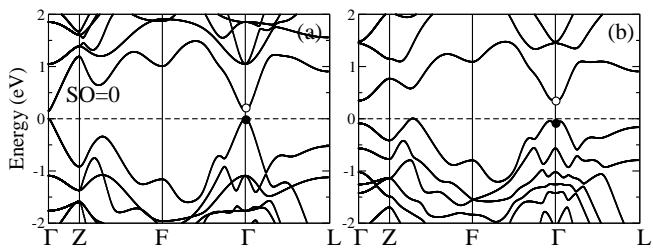}\end{center}
\caption{Bandstructures of  bulk Bi$_2$Se$_3$ (a) without and (b) with
spin-orbit interaction. Dashed lines show the position of the Fermi level, taken
as a reference energy. Filled(open) circles indicate the position of VBM(CBM) at
the $\Gamma$ point of the bulk Brillouin zone. The Brillouin zone of the bulk crystal 
and the path along the high-symmetry directions are depicted in Fig.~\ref{fig0}(c).}\label{fig1} 
\end{figure}

An intuitive way of describing band inversion theoretically is to look at charge
density distributions of the inverted bands at the point of the Brillouin zone,
where the inversion is expected to occur, as functions of a characteristic
parameter.~\cite{Yan} Here we apply this procedure to bulk Bi$_2$Se$_3$. The
calculated bandstructures without and with SO are shown in Fig.~\ref{fig1}(a)
and (b), respectively. The effect of SO is clearly seen as an increase of the
gap and a change in the curvature of the valence band at the $\Gamma$ point.  

\begin{figure}[ht!]
\begin{center}\includegraphics[width=0.98\linewidth,angle=0,clip=true]{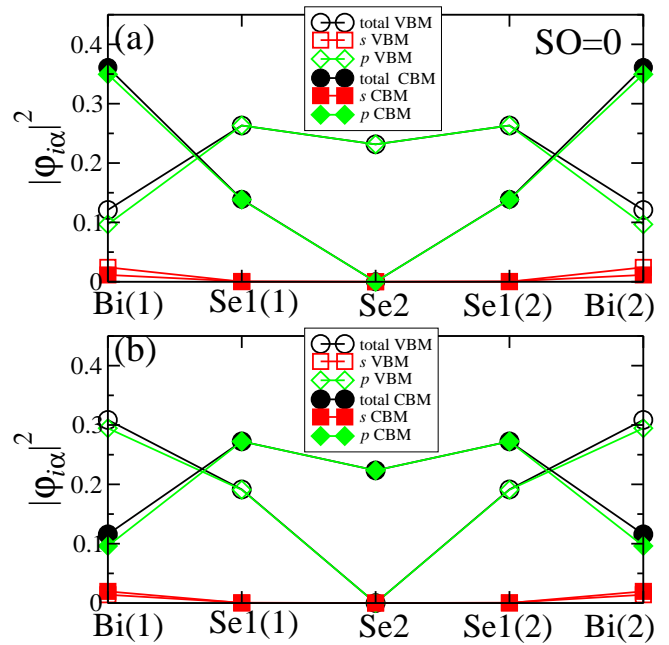}\end{center}
\caption{(Color online) Spatial distribution of the total and orbital-projected
wavefunctions, associated with the valence-band maximum (VBM) and 
the conduction-band minimum (CBM) at the $\Gamma$ point in bulk
Bi$_2$Se$_3$ (a) without and (b) with spin-orbit interaction.}
\label{fig2} 
\end{figure}

To further quantify the changes in the valence and conduction band character, we
calculate the atomic- and orbital-projections of the wavefunctions associated
with the eigenvalues at the valence-band minimum (VBM) 
[filled circles in Fig.~\ref{fig1}], and the conduction-band maximum (CBM) 
[open circles in Fig.~\ref{fig1}] at the $\Gamma$ point. The orbital-resolved
projection of the wavefunction $\left| \varphi \right\rangle$, 
corresponding to an
eigenvalue $\varepsilon_n$ of the TB Hamiltonian in Eq.~(\ref{eq:1}), onto
atomic orbital $\left| i,\alpha,\sigma\right\rangle$ is calculated as
$|\varphi_{i\alpha}|^2\equiv \sum_{\sigma} |\left\langle i,\alpha,\sigma \right|
\left. \varphi \right\rangle|^2$. The total weight of the wavefunction 
at the atomic
site $i$ is then given by $|\varphi_{i}|^2= \sum_{\alpha}
|\varphi_{i\alpha}|^2$. The calculated spatial distribution and orbital
character of the wavefunctions at VBM and CBM are plotted in Fig.~\ref{fig2} for
all five atoms of the bulk unit cell.  

The most noticeable feature of Fig.~\ref{fig2} is the \textit{inversion} in the
spatial distribution of the wavefunctions when SO is switched on.  Without SO
the CBM wavefunction has a node at Se2, which is in fact the inversion center of
the bulk crystal, with the maximum weight on the two Bi atoms.  In contrast, the
maximum weight of the VBM wavefunction is distributed over Se2 and two Se1
atoms. With SO the situation is the opposite: now the VBM wavefunction has a
node at Se2 while the CBM wavefunction is predominantly localized on Se atoms. 

As expected, both VBM and CBM states are predominantly of $p$-character. The
$s$-orbital contribution to the wavefunctions is approximately $10\%$ for
calculations with or without SO and the $s$-projections are not affected by the
inversion. Note that in the presence of SO, the states at the $\Gamma$ point are
mostly originating from $p_z$ orbitals, however the $p_{x(y)}$ (in-plane)
contribution is not negligible ($40\%$ for VBM and $20\%$ for CBM). 

In the next section, we will discuss the results of bandstructure calculations
for Bi$_2$Se$_3$ slabs, namely the emergence of conducting states across the
inverted bulk gap in the presence of SO. In particular, we will show that the
band inversion, characteristic of the bulk gap, can also be identified in the
surface bandstructures.

\subsection{Wavefunction-based analysis of surface 
states in Bi$_2$Se$_3$ slab}\label{slab}

\subsubsection{Surface bandstructures and thickness-dependent gap}\label{gap}

\begin{figure}[ht!]
\begin{center}\includegraphics[width=0.98\linewidth,angle=0,clip=true]{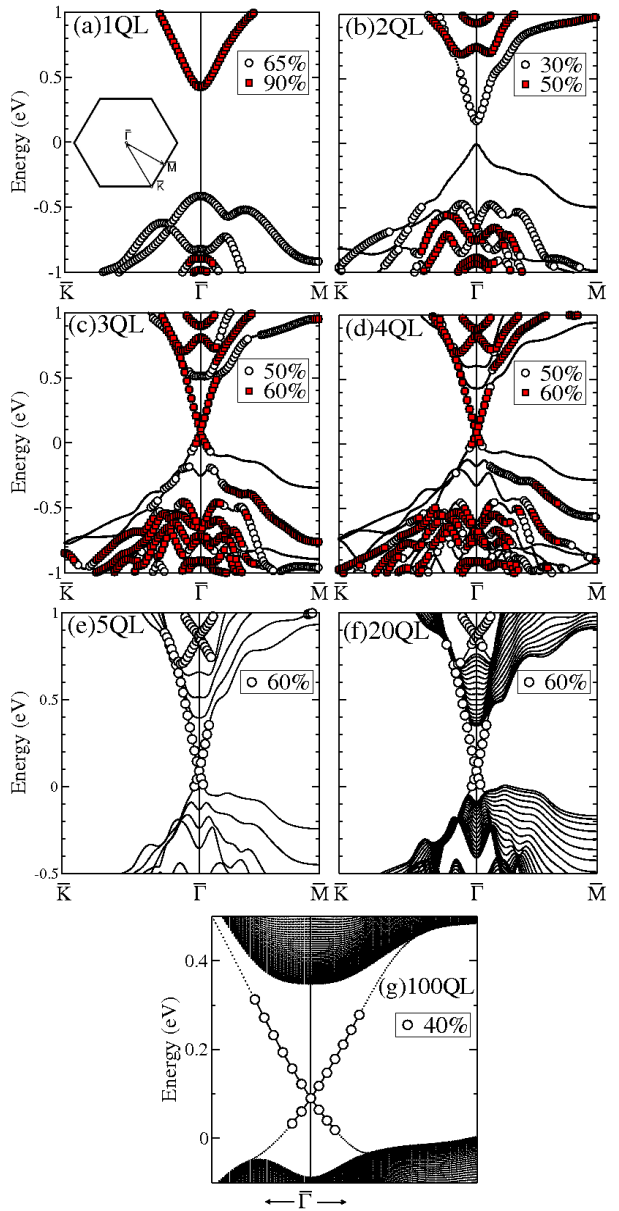}\end{center}
\caption{(Color online) Bandstructures of a  Bi$_2$Se$_3$ slab of varying
thickness. Panels (a)-(e) correspond to 1-5 QLs, (f) 20 QLs and (g) 100 QLs. The inset of panel (a) shows the
two-dimensional Brillouin zone of (111) surface of Bi$_2$Se$_3$. Symbols mark 
the surface states, identified according to the critical weight of the 
wavefunction projection onto the surfaces of the slab (see text for details).}
\label{fig3} 
\end{figure}

The bandstructures of Bi$_2$Se$_3$ slab of varying thickness, calculated using 
the TB model including SO, are shown in Fig.~\ref{fig3}.  A clear and sizable
gap is found for 1 and 2 QLs ($\Delta$=$0.84$~eV and $\Delta$=$0.18$~eV,
respectively). Starting from 3 QLs, two surface bands, resembling the Dirac
states, extend in the range of [-0.1;0.4] eV, with valence and conduction bands
beginning to form below and above this range. Already for 3 and 4 QLs the two
bands appear to be almost touching at the $\bar{\Gamma}$ point, however the
calculated gap is not negligible and is found to be $\Delta$=$0.043$~eV for 3QL
and $\Delta$=$0.007$~eV for 4 QLs. The presence of the gap due to finite
thickness is a recently discovered feature of 3D TI thin films, which has been
attributed to tunneling between surface states localized on the opposite
surfaces of the film.~\cite{YZhang} The size of the gap decreases with
increasing the film thickness and is almost negligible for 6 QLs, which is
quantitatively consistent with our numerical observations. Such gap-opening
mechanism has been considered for possible applications in TI-based MOSFET
devices.~\cite{Chang} 

\begin{figure}[ht!]
\begin{center}\includegraphics[width=0.98\linewidth,angle=0,clip=true]{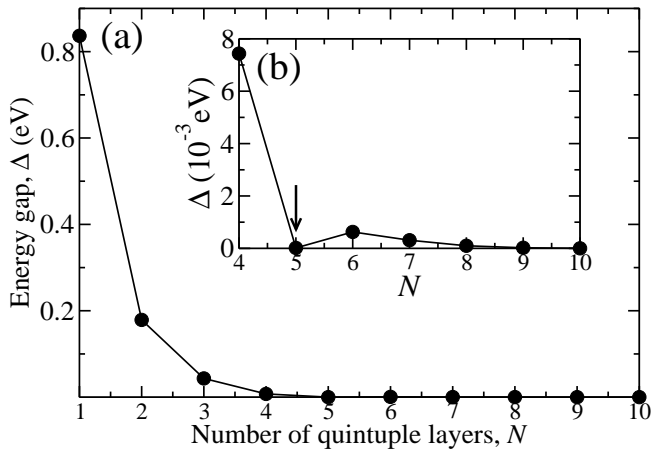}\end{center}
\caption{(a) The energy gap at the 
$\bar{\Gamma}$ point as a function of the slab thickness.  Panel (b) is the zoom
in of (a), showing a local minimum in the thickness dependence of the gap
occurring at 5 QLs.} \label{fig4} 
\end{figure}

The value of the gap for increasing slab thickness is plotted in
Fig.~\ref{fig4}. Interestingly, we find that the gap initially decreases and
reaches the value of $1.7\cdot 10^{-5}$~eV for 5 QLs but then increases up to
$6.2\cdot 10^{-4}$~eV for 6 QLs.  After this sudden increase, the  gap continues
to decrease exponentially and we find $\Delta$=$2.6\cdot 10^{-6}$~eV for 10 QLs.
A similar non-monotonicity in the thickness-dependence of the gap was found in
the range of 4-6 QLs in \textit{ab initio} calculations for Bi$_2$Te$_3$ thin
films.~\cite{KPark} This result also resembles the oscillating behavior of the
gap found in theoretical work based on effective models.~\cite{CXLiu,Linder,Lu2010}
However, direct comparison with these effective-model results 
is not straightforward since, in 
contrast to our atomistic model with slab thickness measured in terms of
elementary building blocks of the real material (QLs), there the gap is
calculated as a continuous function of the size along the $z$-direction.
 
Since the gap decreases exponentially, we do not display the results for
thicknesses greater that 10 QLs. Note, however, that for 40 QLs the size of the
gap is smaller than the numerical accuracy of our calculations. For thicknesses
larger than this critical value, we identify the crossing of the two branches of
Dirac states with opposite group velocity as the Dirac point, which is
found at $0.09$~eV. In Section~\ref{wf_atG} we will investigate in more detail
the effect of the coupling between the two surfaces of the slab on the states at
the Dirac point. In particular, we will look at the spatial character of the
wavefunctions associated with these states and their dependence on the slab
thickness.  

We will now comment on another feature of Fig.~\ref{fig3}, namely on the
procedure used to identify the surface states in the bandstructure calculations.
The character of each Bloch state $\varepsilon_{n}({\bf k})$, where $n$ is the
band index and ${\bf k}$ is momentum, is determined by the spatial distribution
of the corresponding wavefunction:~\cite{KPark} if the relative weight of the
atom-projected wavefunction $\sum_i|\varphi_i|^2$ on the top and bottom QLs
exceeds a critical value (critical percentage) $\gamma$, $\varepsilon_n({\bf
k})$ is identified as a surface state. Such criterion is reasonable for slabs
with thickness greater than 3 QLs, since the typical penetration depth of the
surface states in Bi$_2$Se$_3$ is of the order of 1 QL, or 1~nm.~\cite{WZhang}
For ultra-thin slabs with thickness below 3QLs, the criterion has to be
modified, namely we calculate the wavefunction projections onto top two and
bottom two atomic layers. The value of the critical percentage is found
empirically. Starting from a rough estimate of $\gamma$ based on the ratio
between the penetration depth and the slab thickness, the value should then be
optimized in such a way that a small change around this value does not
significantly modify the resulting distribution of  the surface states. 

We find, however, that for 1-3 QLs the identification of such an optimized value
of $\gamma$ is problematic. To illustrate this point, we present the results of
calculations for two different values of $\gamma$. In the case of 1 QL and
$\gamma$=$65\%$, for instance, all states in the range of energies considered
can be identified as surface states. Hence, for such a thin slab it is
reasonable to use a stronger criterion. As one can see from Fig.~\ref{fig3}(a),
for $\gamma$=$90\%$ only the upper band preserves the surface character while
the lower band does not satisfy the criterion (we use terms "lower" and "upper"
for the two bands defining the gap). Interestingly, this observation is
consistent with scanning tunneling spectroscopy measurements on Sb$_2$Te$_3$
ultra-thin films.~\cite{Jiang} A similar situation is found for 2 QLs, i.e. for
two different values of $\gamma$ the lower band is clearly not a surface band.
For a stronger criterion with $\gamma$=$50\%$ the upper band also partially
looses its surface character. In the case of 3 QLs, varying $\gamma$ in the
range of $50$-$60$\% induces  some visible changes in the distribution of the
surface states, however, the surface character of the Dirac-like states is well
captured for any value of $\gamma$ in this range. For slabs with thicknesses
greater than 3 QLs, the search for an optimized value of $\gamma$ is
significantly simplified. Already for 4 QLs, changing $\gamma$ from $50\%$ to
$60\%$ does not produce any significant difference.  For 5 QLs- and 20 QLs-thick
slabs we use $\gamma$=$60\%$. For a very thick slab of 100 QLs, we focus
particularly on the states in the vicinity of the Dirac point, e.g. in the range
of [0.0;0.3]~eV, and do not apply the procedure for the near-continuum of bulk
states appearing above and below this range.  In Fig.~\ref{fig3}(g) we present
the result for 100 QLs with $\gamma$=$40\%$, where the surface character of the
Dirac states is clearly confirmed. 

Before concluding this section we repeat the procedure that was used in
Section~\ref{bulk} to illustrate the mechanism of band inversion in bulk
Bi$_2$Se$_3$.  For a thick slab [see Fig.~\ref{fig3}(f) and (g)], one can
identify a clear gap between valence and conduction bands, which is transversed
by conducting surface states. As the thickness of the slab increases, the gap
approaches its bulk value, which is found at the $\Gamma$ point in the bulk
bandstructure calculated with SO [Fig.~\ref{fig1}(b)]. VBM(CBM) at the $\bar{\Gamma}$ 
point can be defined 
as the first state below(above) the Dirac point. In the case without SO,  no
conducting surfaces states are present and a gap is found at the $\bar{\Gamma}$
point for all slab thicknesses (calculations are not shown here). Similarly to
the SO case, as the thickness of the slab increases the value of the gap
approaches that found in the bulk calculation without SO [Fig.~\ref{fig1}(a)].
\begin{figure}[ht!]
\begin{center}\includegraphics[width=0.98\linewidth,angle=0,clip=true]{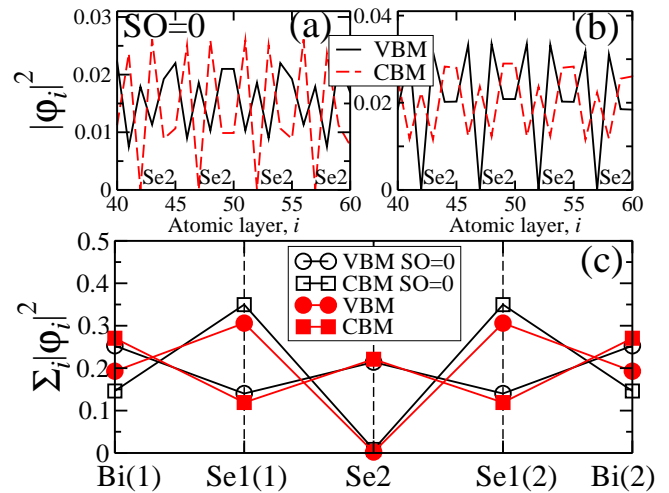}\end{center}
\caption{(Color online) Atomic projections of the wavefunctions, associated with
VBM and CBM at the $\bar{\Gamma}$ point in a 20 QLs-thick slab of Bi$_2$Se$_3$.
Top panels show the atomic-layer projections for a selection of layers in the
middle of the slab (a) without  and (b) with spin-orbit interaction.  $i=1$ is
the first (bottom) layer, $i=100$ is the  last (top) layer. 
(c) The total weights of VBM and CBM wavefunctions 
on five atoms forming a QL (two Se1, Se2 and two Bi atoms), 
summed up of the entire slab.} 
\label{fig5} 
\end{figure}

The atomic-layer projected wavefunctions at VBM and CBM of a 20 QLs-thick slab
without and with SO are shown in Fig.~\ref{fig5}(a) and (b), respectively. One
immediately notices that the spatial character of VBM and CBM is inverted when
SO is switched on: similarly to the bulk case, with(without) SO the VBM(CBM)
wavefunction has nodes on Se2 atoms. The effect becomes even more appreciable,
if we sum the wavefunction projections on Se1, Se2 and Bi atoms over the entire
slab [Fig.~\ref{fig5}(c)]. The result is clearly similar to what we found in the
bulk case, indicating that the inverted character of the valence and conduction
bands in the presence of SO is also an intrinsic property of the surface, with
the difference that in the surface bandstructures one finds the topological
surface states, dispersing linearly across the inverted band gap.  

\subsubsection{Probing the wavefunction at the Dirac point: effect of finite
thickness}\label{wf_atG}

We will now consider the surface  
states and their corresponding wavefunctions, 
found at the $\bar{\Gamma}$ point in our surface bandstructure calculations. In 
a hypothetical situation, when the slab is thick enough so that the two surfaces 
do not interact with each other, one expects to find four degenerate states at 
the $\bar{\Gamma}$ point, namely one Kramers 
degenerate pair for each of the
two surfaces. Away from the $\bar{\Gamma}$ point, 
the four-fold degeneracy is lifted
and one expects a doubly degenerate state at each momentum ${\bf k}$,  
with degeneracy guaranteed by inversion symmetry. 
In addition, for each state at
${\bf k}$, there is an identical state with opposite spin at $-{\bf k}$ due
to time-reversal symmetry.  
\begin{figure}[ht!]
\begin{center}\includegraphics[width=0.98\linewidth,angle=0,clip=true]{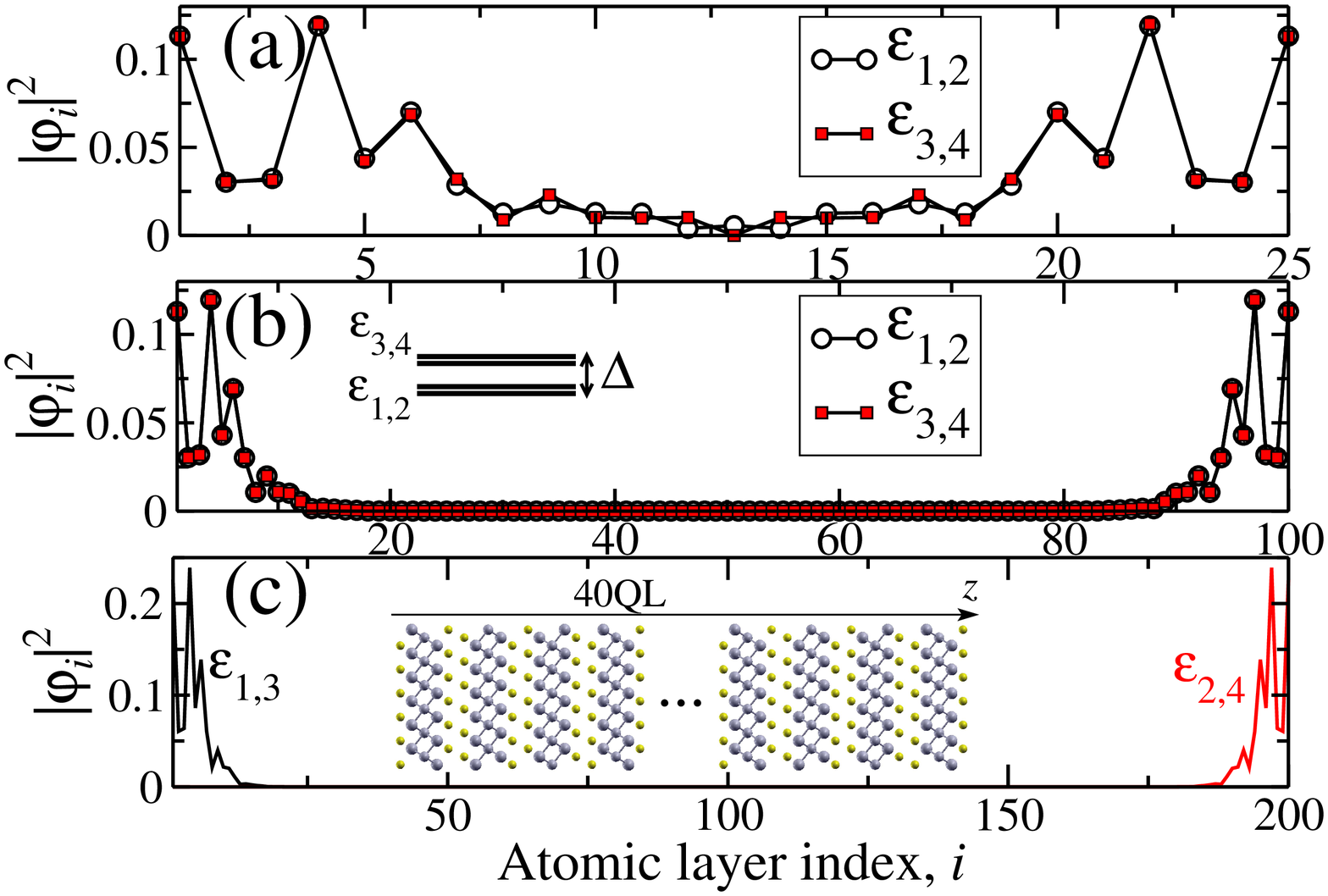}\end{center}
\caption{(Color online) Atomic-layer projections of the wavefunctions,
corresponding to four quasi-degenerate states exactly at the $\bar{\Gamma}$
point for (a) 5 QLs, (b) 20 QLs and (c) 40 QLs of Bi$_2$Se$_3$. The inset of
panel (b) shows schematically the two pairs of degenerate states at
$\bar{\Gamma}$, separated by an energy gap $\Delta$. The inset of panel (c)
displays the crystal structure of a 40 QLs-thick slab of Bi$_2$Se$_3$ with top
and bottom surfaces.} \label{fig6}
\end{figure}

The analysis of the thickness-dependent gap, carried out in the previous
section, shows that the limit of an infinitely thick slab with zero interaction
between the two surfaces is most likely realized for 40 QLs in our system. 
Although the size of the gap already for 6 QLs is considerably smaller than the
value found in ultra-thin slabs, for 40 QLs the gap becomes identically zero
within the numerical precision of our calculations (note that we employ exact
diagonalization to calculate the eigen-spectrum of the Hamiltonian at each ${\bf
k}$). By looking closely at the $\bar{\Gamma}$ point, we indeed find four
degenerate states $\varepsilon_i$ ($i$=1,..,4) in the case of 40 QLs, as shown
in Fig.~\ref{fig6}(c). Within each pair of degenerate states,
$\varepsilon_{1,3}$ and $\varepsilon_{2,4}$, the states have opposite spins and
each pair is localized on either top or bottom surface. Note that the $a$-th
Cartesian component of the spin of each Bloch state is calculated as
$s^{a}_{\varepsilon_{n}}({\bf k})$=$\mathrm{Tr}[\rho_{nn}{\sigma^{a}}]$, where
${\bf \sigma}$=$\left\{\sigma^{a}\right\}$ is the set of Pauli matrices and
$\rho_{nn}$ is the $n$-th diagonal element of the density matrix constructed
from the eigenfunctions of the Hamiltonian [Eq.~(\ref{eq:1})] at momentum ${\bf
k}$. From Fig.~\ref{fig6}(c) we can also estimate the decay length of the
surface states to be approximately 10 atomic layers, or, equivalently, 2 QLs in
agreement with previous reports.~\cite{WZhang, KPark} 
\begin{figure}[ht!]
\begin{center}\includegraphics[width=0.98\linewidth,angle=0,clip=true]{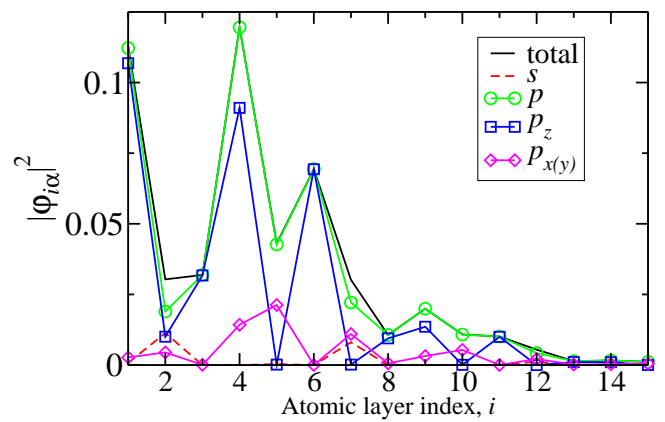}\end{center}
\caption{(Color online) Orbital-resolved atomic-layer projection of the
wavefunction, corresponding to one of the four degenerate states exactly at the
$\bar{\Gamma}$ point for a  20 QLs -thick slab of Bi$_2$Se$_3$. Since the
wavefunction is symmetric with respect to the center of the slab [see
Fig.~\ref{fig6}(b)], only atomic layers belonging to the bottom surface are
 displayed. } \label{fig7} 
\end{figure}

For slabs with thicknesses below 40 QLs, we also find four states at the
$\bar{\Gamma}$ point, however there is a finite gap between degenerate states
$\varepsilon_{1,2}$ and $\varepsilon_{3,4}$ [see the inset in
Fig.~\ref{fig6}(b)]. Within each pair of states, the Bloch states with identical
energy have opposite spins but, in contrast to the 40 QLs case, their
wavefunctions are distributed over both top and bottom sides of the slab,
signaling a non-negligible interaction between the the two surfaces. For 20 QLs,
since the difference in energy between $\varepsilon_{1,2}$ and
$\varepsilon_{3,4}$ is finite but small (less than $10^{-6}$~eV), the spatial
distributions of the corresponding wavefunctions are nearly identical and are
therefore indistinguishable on the scale of the graph [Fig.~\ref{fig6}(b)]. In
the case of 5 QLs, when the gap at the $\bar{\Gamma}$ point is more appreciable,
there  is a clear difference between the spatial profiles of the wavefunctions,
associated with the two degenerate states [Fig.~\ref{fig6}(a)]. The difference
is the largest in the middle of the slab, where the tails of the wavefunctions
of the surface states residing on the opposite sides overlap, creating a mixed
state. 

Finally, we comment on the orbital character of surface states at the Dirac
point. The orbital-resolved atomic-layer projections of the wavefunctions
associated with one of the four quasi-degenerate states ($\varepsilon_1$) at the
$\bar{\Gamma}$ point in a 20 QLs-thick slab are plotted in Fig.~\ref{fig7}.
Similarly to the VBM and CBM states at the $\Gamma$ point  of the bulk
bandstructure with SO [Fig.~\ref{fig1}(b)], this state is predominantly of
$p$-character with $s$-orbital contribution less than $10\%$ percent. Although
the contribution of $p_z$-orbitals is the largest, the relative weight of
in-plane ($p_{x(y)}$) orbitals to the orbital-resolved wavefunction is
non-negligible and is of the order of $40\%$.  The importance of the SO-induced
in-plane orbital contribution to the wavefunction of the Dirac states in
Bi$_2$Se$_3$ has been recently demonstrated using orbital-selective SR-ARPES
measurements.~\cite{Cao}

\subsubsection{Spin-resolved surface states on the Dirac cone}\label{spin}

In this section we analyze the surface states away from the $\bar{\Gamma}$
point. We describe the spatial distribution and the spin-properties 
of the wavefunctions, 
associated with energy states found inside the inverted band gap at non-zero
values of momentum ${\bf k}$. In particular, we quantify the
effect of the finite slab thickness on the helical spin-character of the Dirac
cone states.  
\begin{figure}[ht!]
\begin{center}\includegraphics[width=0.98\linewidth,angle=0,clip=true]{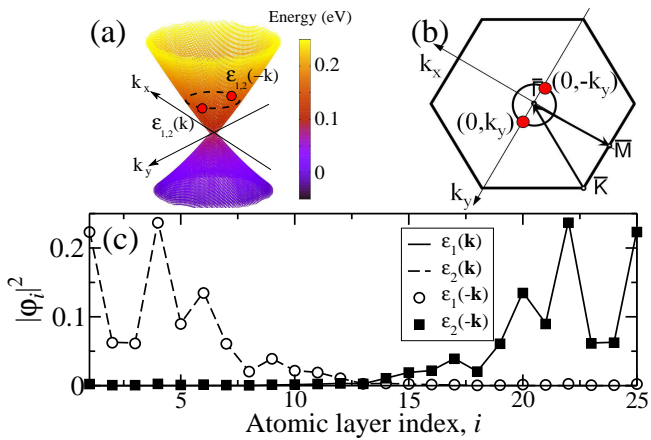}\end{center}
\caption{(Color online) (a) Dirac cone calculated for a 5 QLs-thick slab,  in
the energy range of [-0.05:0.25]~eV. An equi-energy contour above the Dirac
point (at approximately 0.13~eV) is shown with a dashed line. Red circles
indicate a doubly degenerate state $\varepsilon_{1,2}$ occurring at $\bf{k}$ and
its time-reversal partner at $-\bf{k}$. (b) Schematic of the two-dimensional
Brillouin zone with the projection of the equi-energy contour and two doubly
degenerate states at $\pm\bf{k}$=$\{0,\pm k_y\}$. (c) Atomic-layer projections
of the wavefunctions associated with four states
$\varepsilon_{1,2}(\pm\bf{k})$.} \label{fig8} 
\end{figure}

Figure~\ref{fig8}(a) shows an equi-energy contour located slightly above the
Dirac point, within the energy range containing the Dirac cone, for a 5QLs-thick
slab. There is a doubly degenerate state  at each ${\bf k}$ on the contour and a
corresponding doubly degenerate state at $-{\bf k}$  with the same energy. We
first consider a pair of doubly degenerate states $\varepsilon_{1,2}({\bf k})$
and $\varepsilon_{1,2}(-{\bf k})$ with ${\bf k}=(0,k_y)$ as shown in
Fig.~\ref{fig8}(b). The atomic-layer projections of the wavefunction for all
four states are presented is Fig.~\ref{fig8}(c). Each of the degenerate states
within a pair $\varepsilon_{1,2}(\pm{\bf k})$ is localized on either top or
bottom surface. We will refer to the state whose wavefunction is localized
predominantly on the top (bottom) surface as a top (bottom) state.

\begin{figure}[ht!]
\begin{center}\includegraphics[width=0.98\linewidth,angle=0,clip=true]{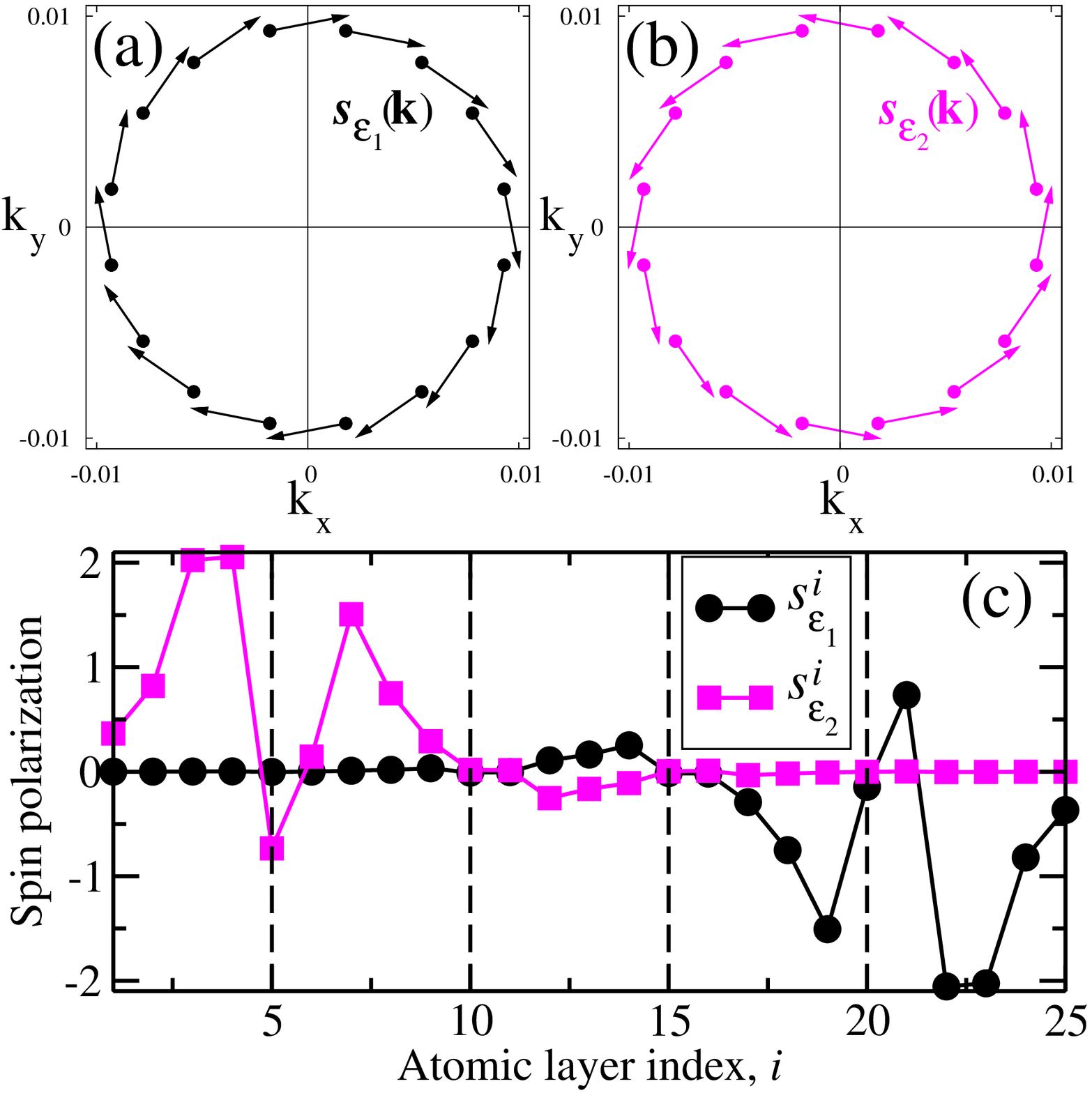}\end{center}
\caption{(Color online) Spin-momentum locking: the in-plane projection of the
spin, associated with degenerate states (a) $\varepsilon_1$  and (b)
$\varepsilon_2$ for several values of momentum ${\bf k}$ on the equi-energy
contour [see Fig.~\ref{fig8}(a)].} \label{fig9} 
\end{figure}

Furthermore, we calculate the spins of top and bottom states at several momenta
on the equi-energy contour. Figures~\ref{fig9}(a) and (b) show the projection of
the spin on the $x$-$y$ plane in momentum space for top and bottom states,
respectively. The direction of the spin at each ${\bf k}$ appears to be
tangential to the equi-energy contour, which is a manifestation of the
spin-momentum locking intrinsic to 3D TIs.~\cite{Hsieh2009} The direction of the
spin is exactly opposite for the states $\varepsilon_1({\bf k})$ and
$\varepsilon_2({\bf k})$, hence the top and bottom surface states have opposite
helicities. These observations indicate that despite the opening of the gap at
the $\bar{\Gamma}$ point due to interaction between the two surfaces for a
5QLs-thick slab, the states appearing across the bulk band gap at this thickness
are of topological character.  

\begin{figure}[ht!]
\begin{center}\includegraphics[width=0.98\linewidth,angle=0,clip=true]{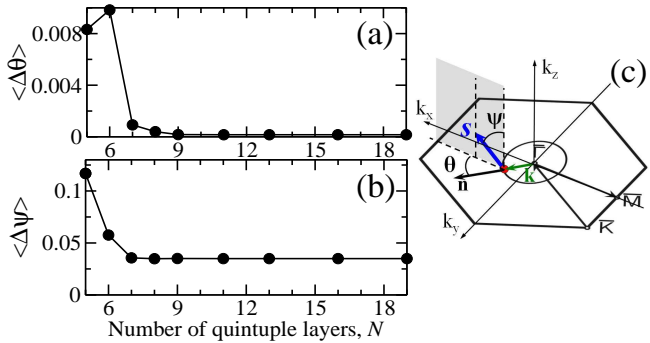}\end{center}
\caption{(Color online) Deviation of the spin (${\bf s}$), associated with one
of the states  $\varepsilon_{1,2}$ on the equi-energy contour, (a) from
direction perpendicular to $\bf{k}$  and (b) from in-plane orientation. (c)
Orientation of ${\bf s}$ in momentum space: $\theta$ is the angle between ${\bf
s}$ and the normal ${\bf n}$ to the equi-energy contour, $\psi$ is the angle
between ${\bf s}$ and the direction perpendicular to the plane. We define
$\Delta\theta(\Delta\psi)$=$|90^{\circ}-\theta(\psi)|$; $\left\langle
\Delta\theta(\Delta\psi)\right\rangle$ is the average of
$\Delta\theta(\Delta\psi)$ over ${\bf k}$ on the equi-energy contour.}
\label{fig10}
\end{figure}

In order to investigate the effect of the slab thickness on the helical
spin-texture of the surface states, we determine the orientation of the spin
${\bf s}$ of the Bloch states $\varepsilon_{1,2}({\bf k})$ for thicknesses in
the range of 5 to 20 QLs. The orientation of vector ${\bf s}$ in momentum space
at each ${\bf k}$ can be described by two angles: $\theta$, or in-plane tilting
angle, which is the angle between ${\bf s}$ and the normal to the contour at
point ${\bf k}$, and $\psi$, or out-of-plane tilting angle, which in the angle
between ${\bf s}$ and the direction perpendicular to the plane [see
Fig.~\ref{fig10}(c)]. For a perfect spin-momentum locking
$\theta$=$\psi$=$90^{\circ}$. Note that the angles $\theta$ and $\psi$ oscillate
as functions of ${\bf k}$ along the equi-energy contour due to the three-fold
rotational symmetry of the slab crystal structure.~\cite{Zhao} In addition, at
each value of momentum the in-plane and out-of-plane tilting angles of the top
and bottom states are equal in magnitude and have opposite signs. Hence we
compute the deviations of $\theta$ and $\psi$ from $90^{\circ}$
only for the top state. The resulting absolute-value deviations, averaged 
over $\bf k$ on the equi-energy contour,
$\left\langle \Delta\theta\right\rangle$ and 
$\left\langle \Delta\psi\right\rangle$, are
plotted   as functions of the thickness in Figs.~\ref{fig10}(a) and (b),
respectively.  

Both in-plane and out-of-plane deviations decrease with increasing the slab
thickness. The dependence of $\left\langle \Delta\theta\right\rangle$ shows a
kink at 6 QLs, consistent with the increase of the gap at the $\bar{\Gamma}$
point (see Fig.~\ref{fig4}), while $\left\langle \Delta\psi\right\rangle$
decreases monotonically.  Importantly, both $\left\langle
\Delta\theta\right\rangle$ and $\left\langle \Delta\psi\right\rangle$ do not
decrease to zero with increasing the thickness but instead saturate to a
constant value. The saturation  value is small for $\left\langle
\Delta\theta\right\rangle$ (${2\cdot 10^{-4}}^{\circ}$) and relatively large for
$\left\langle \Delta\psi\right\rangle$ ($0.03^{\circ}$).  The saturation starts
at approximately 10 QLs for $\left\langle \Delta\theta\right\rangle$ and at 7
QLs for $\left\langle \Delta\psi\right\rangle$.  The non-zero residual deviation
from the perfect helical spin-texture at large thicknesses is a measure of
the proximity of the equi-energy contour to bulk states. 
The further the contour is 
from the Dirac point, the stronger is the effect of the bulk continuum, with
hexagonal wrapping in momentum space due to crystal symmetry, on the surface
states located on the contour. For an equi-energy contour located at a higher
energy ($0.26$~eV) we find considerably larger saturation values, namely
$0.5^{\circ}$ for $\left\langle \Delta\psi\right\rangle$ and $0.1^{\circ}$ for
$\left\langle \Delta\theta\right\rangle$. To summarize, both in-plane and out-of
plane deviations decrease with increasing the slab thickness and with moving the
equi-energy contour closer to the Dirac point. However in the vicinity of the
Dirac point the in-plane tilting angle appears to be more affected by the finite
thickness while the out-of-plane tilting angle is more sensitive to the presence
of bulk states.
    
\section{Conclusions}\label{concl}

We have employed the \textit{sp}$^3$ tight-binding model, with parameters
extracted from \textit{ab initio} calculations, to model the electronic
structure of bulk and (111) surface of Bi$_2$Se$_3$ 3D TI. We presented a
quantitative description of the band inversion mechanism for both bulk and slab
geometry, which involves a detectable change in the spatial distribution of
$p$-orbital projections of the wavefunctions of conduction and valence bands
induced by spin-orbit interaction. The surface bandstructures, with the spatial
character of Bloch states determined by quantitative criteria, were
calculated for slabs with thicknesses up to 100 QLs. This thickness is well beyond what
is accessible by \textit{ab initio} approaches. Based of the calculated
thickness-dependent gap due to inter-surface interaction and the
atomic-projections of the wavefunctions for states at the $\bar{\Gamma}$ point,
we found that the infinite-thickness limit, characterized by zero gap and no
interaction between the surfaces, is realized for 40 QLs.  Furthermore, our
calculations showed that the disturbances in the helical spin-texture of the
Dirac-cone states, caused by the finite slab thickness and the proximity to bulk
states, persist for thicknesses up to 10 QLs even in the vicinity of the Dirac
point.  

We would like to comment on our numerical observation that the top and bottom
surface states become completely decoupled only for thick slabs containing 40
QLs. Strictly speaking only states with linear dispersion and identically zero
gap at the Dirac point can be identified as Dirac states. However, in practice our
observation does not contradict the commonly accepted results that 6
QLs is the critical thickness which determines the topological character of the
material.~\cite{YZhang} Indeed, we found that the gap for thicknesses greater
that 5 QLs is very small (of the order of $10^{-4}$~eV and smaller) and the
surface states appearing inside the bulk insulating gap at these thicknesses
have nearly linear dispersion and helical spin-texture. Nevertheless, the
presence of a weak but finite interaction between the opposite surfaces 
for slabs of less than 40 QLs thick might be
crucial for the analysis of subtle effects in the vicinity of the Dirac point.
Preliminary calculations have shown that even a small perturbation preserving
the time-reversal symmetry can open a gap for thicknesses as large as 20
QLs.~\cite{Pertsova} This may further hinder the identification of the gap
induced purely by the presence of time-reversal-breaking perturbations, such as
magnetic impurities,~\cite{QLiu} in calculations based on tight-binding models
similar to ours and certainly in \textit{ab initio} studies, where thicknesses
are limited to only few QLs. 

Effects due to finite thickness can be also expected in Landau levels
spectroscopy of 3D TI thin films.~\cite{Hanaguri} On one hand, it has been
shown experimentally that in Bi$_2$Se$_3$-like materials the characteristic
field-independent (zero-th) Landau level is absent for slabs with thicknesses
smaller than 3 QLs but emerges already for 4 QLs.~\cite{Jiang} On the other
hand, a recent theoretical study suggests a splitting of the zero-th Landau level
due to hybridization between top and bottom surface states.~\cite{Yang} Based on
these considerations, one might expect that a non-negligible inter-surface
interaction can lead to more subtle internal structure of the zero-th Landau
level, persisting even for relatively large thicknesses.  

We anticipate that microscopic tight-binding models, combined with input from
\textit{ab initio} calculations, will play an increasingly important role in
practical calculations of various properties of TI materials. These include the
detailed character of the surfaces states wavefunction, which to some extent can
be already probed experimentally,~\cite{Cao} and the interplay between the
surfaces states and external perturbations.~\cite{Wray} In fact, a
finite-cluster tight-binding approach, based on the model used in the present
work, has been already employed in the study of native defects in Bi$_2$Se$_3$,
showing good agreement between the calculated local densities of states around
the defects and experimental STM topographies.~\cite{Mahani} In connection to
this last point, we would like to mention that despite extensive studies of the
effect of magnetic doping in 3D TIs, a consistent microscopic description of a
single magnetic impurity in a TI environment appears to be incomplete,
especially when compared to the progress that has been made in investigating similar questions 
in semiconductors, both theoretically~\cite{Tang,Strandberg} and
experimentally.\cite{Yakunin, Kitchen} A realistic tight-binding approach can be
indispensable in providing such a microscopic description for single impurities
in 3D TIs. 

\begin{acknowledgments}
 
We are grateful to A.~H.~MacDonald for illuminating discussions. 
We acknowledge helpful interactions with M.~R.~Mahani. This work was
supported by the Faculty of Natural Sciences at Linnaeus University and by the
Swedish Research Council under Grant Number: 621-2010-3761. Computational resources have
been provided by the Lunarc center for scientific and technical computing at
Lund University. 

\end{acknowledgments}

\end{document}